\documentclass[submission, Phys]{SciPost}

\usepackage{amssymb}
\usepackage{mathtools}
\usepackage{bm}
\usepackage[section]{placeins}

\newcommand{\dvec}{\mathbf d}
\newcommand{\mvec}{\mathbf m}
\newcommand{\nvec}{\mathbf n}
\newcommand{\rvec}{\mathbf r}

\newcommand{\sigmavec}{\boldsymbol{\sigma}}
\newcommand{\QR}{Q_R}
\newcommand{\QB}{Q_B}
\newcommand{\zR}{z_R}
\newcommand{\zL}{z_L}
\newcommand{\overlap}{\langle z_L|z_R\rangle}
\newcommand{\Hzeff}{\hat e_z}
\newcommand{\Heff}{H_{\mathrm{eff}}}
\newcommand{\Sigmacom}{\Sigma}
\newcommand{\PTsym}{\mathcal{PT}}
\newcommand{\Ecal}{\mathcal{E}}
\newcommand{\tvec}{\mathbf t}

\begin{document}

\begin{center}{\Large \textbf{
A real-space exceptional ring mediates an eigenframe-charge transition in a non-Hermitian skyrmion
}}\end{center}

\begin{center}
Kejun Liu\textsuperscript{1,2*}
\end{center}

\begin{center}
{\bf 1} State Key Laboratory of Bioinspired Interface Material Science, Institute of Nano \& Functional Materials, Soochow University, Suzhou 215123, China
\\
{\bf 2} School of Physical Science and Technology, Soochow University, Suzhou 215006, China
\\
* kjliu@suda.edu.cn
\end{center}

\begin{center}
\today
\end{center}

\section*{Abstract}
{\bf
The integer topological charge of a skyrmion is the standard example of topological protection. We
ask what happens to that protection when the local generator is made non-Hermitian by
polarization-selective gain or loss. The texture charge of a smoothly evolving state remains
homotopy-protected under the usual fixed-boundary condition: after projective normalization, the
gain/loss contribution has the Gilbert relaxation form on the target sphere. The spectral
charges of the local generator behave differently. Its biorthogonal eigenframe charge is quantized
while the texture avoids exceptional points. At linewidth matching, the skyrmion equator becomes
a real-space exceptional ring on which the biorthogonal Bloch field diverges, and
the eigenframe class jumps from $+1$ to $0$ in a parameter-driven spectral transition. The closed
contour relies on constant splitting along the equator, or an equivalent design constraint; a
partial or amplitude-anisotropic realization instead leaves isolated real-space exceptional
points. We use an operational Jones parametrization of an isolated polarization doublet and give a
Stokes-tomography reconstruction of the left--right frame. The result identifies a
spatially resolved exceptional contour that mediates an integer transition of a real-space
eigenframe charge.
}

\section{Introduction}
\label{sec:intro}

Topological protection is one of the organizing ideas of condensed-matter physics: an integer
invariant fixed by the global structure of a field cannot change under a smooth local
perturbation. A skyrmion is the canonical real-space example. Its winding number
$Q\in\pi_2(S^2)=\mathbb{Z}$ underlies its stability~\cite{skyrme1962,nagaosa2013}. The same
homotopy appears in polarization textures of light~\cite{tsesses2018,shen2024} and in spinor
polariton microcavities~\cite{cilibrizzi2016}.

Open photonic and polaritonic systems add a different structure. Their polarization generators are
non-Hermitian, their left and right eigenstates differ, and exceptional points (EPs) occur where
eigenvalues and eigenvectors coalesce~\cite{heiss2012,ashida2020,bergholtz2021}. Non-Hermitian band
topology is by now mature~\cite{shen2018,kawabata2019,gong2018}, but it usually concerns Bloch
eigenstates in momentum space. Here we ask what happens to the charge of a \emph{real-space}
eigenstate texture when its local polarization generator reaches an EP.

Nearby work makes the distinction important. Polarization-dependent loss, Voigt EPs, and Stokes
textures have been measured in an anisotropic planar microcavity~\cite{richter2019}. In a separate
theoretical and numerical study, confined optical-skyrmion modes were taken through a global modal
EP using circularly polarized gain and loss~\cite{krol2023}. A phenomenological polariton model
also hosts polarization-dependent linewidths and momentum-space EPs~\cite{hu2025}.
Reference~\cite{krol2023} resolves real-space Stokes textures in its calculated modes and discusses
their pixel-wise polarization evolution, but its EP is the coalescence of one pair of global
confined-mode eigenvalues. Our question is different. At one control value only the local
discriminant on the texture's equator vanishes, while the rest of the same texture remains
nondefective. We then compute the eigenstate-line Chern class on both sides of that real-space
exceptional contour.

We find three levels of behavior, involving distinct mathematical objects rather than three
measurements along one dynamical sweep. First, the charge of a smoothly evolving state is
homotopy-protected when the normalized texture stays continuous and its value at infinity is fixed.
Second, a control parameter drives the instantaneous eigenframe from class $+1$ to class $0$
through a real-space exceptional ring; this is a spectral transition, not the time evolution of
one state. Third, the geometry and coverage of the differential loss determine whether the
defective set is a closed contour or a set of isolated points. A closed contour supports a global
integer transition on its two sides for the family studied here. An isolated EP pair instead
obstructs a global band label, so a continuation integral through that texture is not a topological
charge.

We first define the texture and eigenframe charges (Sec.~\ref{sec:charges}) and establish the
homotopy protection of the evolving state (Sec.~\ref{sec:right}). We then derive the real-space
exceptional ring and its eigenframe-charge transition (Sec.~\ref{sec:ring}), determine when the
ring deforms or splits (Sec.~\ref{sec:geometry}), and study localized differential loss
(Sec.~\ref{sec:localized}). Section~\ref{sec:protocol} gives the polarization protocol and its
domain of validity, followed by the discussion in Sec.~\ref{sec:discussion}.

\section{Texture and eigenframe charges}
\label{sec:charges}

We give explicit formulas for a CP$^1$ spinor $z(\rvec)\in\mathbb{C}^2$, $z^\dagger z=1$, defined
up to a local phase. In the Hermitian theory its real Bloch vector is
\[
n_a(\rvec) = \hat z^\dagger \sigma_a \hat z , \qquad \hat z = z/|z|,
\]
which lives on the unit sphere $S^2$, and the topological charge is the degree of the map
$\rvec\mapsto n(\rvec)$,
\begin{equation}
\label{eq:QR}
\QR = \frac{1}{4\pi}\int \nvec\cdot(\partial_x \nvec\times\partial_y \nvec)\,\mathrm{d}^2r \in \mathbb{Z}.
\end{equation}
For a smoothly evolving state this is the ordinary \emph{texture charge} $Q$. Applied instead to
the right eigenstate of a local generator, the same formula defines the \emph{right-eigenframe
charge} $\QR$.

When the generator of the dynamics is not Hermitian, $H\neq H^\dagger$, a single state vector no
longer captures the physics. The eigenproblem has distinct right and left eigenvectors,
$H|\zR\rangle = E|\zR\rangle$ and $\langle \zL|H = E\langle \zL|$. A corresponding spectral
texture variable is the \emph{biorthogonal Bloch vector}
\begin{equation}
\label{eq:m}
m_a(\rvec) = \frac{\langle \zL|\sigma_a|\zR\rangle}{\overlap},
\end{equation}
which reduces to $n_a$ when $\zL=\zR$ but is in general complex. Biorthogonal averages of this
form enter non-Hermitian spectral geometry and response~\cite{kunst2018,edvardsson2020,mandal2024}.
Pauli completeness gives
$m\cdot m = 1$, so $m$ lives not on the real sphere but on the complex quadric
$\Sigmacom=\{\,m\in\mathbb{C}^3 : m\cdot m=1\,\}$.

We define the \emph{biorthogonal eigenframe charge} $\QB$ by Eq.~(\ref{eq:QR}) with $\mvec$ in place
of $\nvec$. The two-form
$\Omega=(8\pi)^{-1}\epsilon_{abc}m_a\,\mathrm d m_b\wedge\mathrm d m_c$ is closed on $\Sigmacom$, and
$\Sigmacom\cong T^*S^2$ deformation-retracts to its real zero section. Thus any texture that
extends continuously to the compactified plane and has a nonsingular biorthogonal frame carries an
integer $\QB$~(App.~\ref{app:twolevel}). On an EP-free texture it is the first Chern number of the
same right-eigenstate line bundle measured by $\QR$, hence $\QB=\QR$ as an integer even though their
local fields and charge densities differ. Quantization is conditional, not gradually lost. It can
change only when the eigenstate projector ceases to define a continuous line bundle. In the
exceptional-point family studied below this failure appears as vanishing phase rigidity and an
unbounded spectral projector. Thus $\QR$ and $\QB$ are not independent topological classifications;
the specifically biorthogonal information lies in the local field $\mvec$, its charge density, and
the phase-rigidity singularity.

\section{Homotopy protection of the evolving state}
\label{sec:right}

Consider a local CP$^1$ generator with an anti-Hermitian, balanced gain/loss piece along the easy
axis,
\begin{equation}
\label{eq:flow}
i\,\partial_t z = (H_0 + i\gamma\,\sigma_z)\,z ,
\end{equation}
with $\gamma$ the strength of the non-Hermiticity. Equation~(\ref{eq:flow}) does not preserve the
norm of $z$; the physical, projective degree of freedom is the normalized spinor $\hat z=z/|z|$, and
one should ask how the right Bloch vector $n_a=\hat z^\dagger\sigma_a\hat z$ evolves. Write
$H_0=b_0\mathbb 1+\mathbf b\cdot\sigmavec$, $P=\hat z\hat z^\dagger$, and the Hermitian gain/loss
coefficient $\Gamma=\gamma\sigma_z$. The projective evolution is
\begin{equation}
\label{eq:dotP}
\dot P = -i[H_0,P]+\{\Gamma,P\} - 2\,\mathrm{Tr}(\Gamma P)\,P ,
\end{equation}
the last term preserving $\mathrm{Tr}\,P=1$. Taking $n_a=\mathrm{Tr}(\sigma_a P)$ and reducing
Eq.~(\ref{eq:dotP}) gives the closed-form flow
\begin{equation}
\label{eq:gilbert}
\dot{\nvec} = 2\mathbf b\times\nvec
+2\gamma\,\bigl(\Hzeff - n_z\,\nvec\bigr), \qquad \nvec\cdot\dot{\nvec} = 0 .
\end{equation}
We have verified Eq.~(\ref{eq:gilbert}) by symbolic computation; the derivation is recorded in
App.~\ref{app:gilbert}. Both terms are tangent to $S^2$. The first is the usual Hermitian precession.
The gain/loss contribution, namely the second term, has the Gilbert relaxation form
$\Heff-(\nvec\cdot\Heff)\nvec$, with effective field $\Heff=2\gamma\,\Hzeff$ and the conventional
damping normalization absorbed into $\Heff$. The gain/loss
deformation therefore adds a Gilbert-type relaxation toward the $\Hzeff$ axis without taking the
normalized right state off the sphere.

The consequence for the charge follows under the standard compactification hypothesis. If
$\nvec(\rvec,t)$ is jointly continuous, $z$ remains nonzero, and the value at spatial infinity is
fixed, then the evolution is a homotopy of maps $S^2_{\rm space}\to S^2$ and the
degree~(\ref{eq:QR}) is unchanged. A charge change requires a zero of $z$, loss of continuity, or
topological flux through a boundary. This is the structural reason why Hermitian precession and the
additional Gilbert-type term both preserve the texture charge under these conditions. In
Sec.~\ref{sec:ring}, by contrast, we study the
instantaneous eigenstate texture of a local non-Hermitian generator. Across an exceptional point
that eigenstate branch itself is not smooth, so the homotopy argument of the present paragraph does
not apply to it.

This conservation law concerns the evolving state. It does not constrain an instantaneous
eigenframe, whose band can become singular at an EP. That is the question of the next section.

\section{Real-space exceptional ring and eigenframe-charge transition}
\label{sec:ring}

To probe the biorthogonal charge we place the non-Hermiticity in a local generator acting on the
texture. Let $\dvec(\rvec)=h(\sin f\cos\theta,\sin f\sin\theta,\cos f)$ be a unit-winding hedgehog
field of constant modulus $h$. With the spatial orientation used in Eq.~(\ref{eq:QR}),
$\dvec/h$ has degree $-1$. We follow the lower band, whose Hermitian Bloch vector is
$-\dvec/h$ and therefore has $\QR=+1$. Reversing the spatial orientation flips the nonzero plateau
but not the transition to zero. The local generator is
\begin{equation}
\label{eq:Hloc}
H(\rvec) = \dvec(\rvec)\cdot\sigmavec + i\gamma\,\sigma_z ,
\end{equation}
the Hermitian part orienting the pseudospin along the texture and the $i\gamma\sigma_z$ piece
supplying balanced gain and loss.

We parametrize the local response of an isolated polarization doublet, without identifying
momentum with position. Its Jones or coupled-mode generator is
\begin{equation}
\label{eq:Jones}
H_J(\rvec)=[\bar\omega(\rvec)-i\bar\Gamma(\rvec)]\mathbb 1
+h(\rvec)\mathbf S_0(\rvec)\cdot\sigmavec
+i\delta\Gamma(\rvec)\hat{\mathbf e}_\Gamma\cdot\sigmavec .
\end{equation}
Here $\mathbf S_0$ is the eigenpolarization Stokes texture measured without differential loss,
$2h$ is the local polarization splitting, and $\delta\Gamma$ is the differential-linewidth
coefficient along a fixed polarization axis $\hat{\mathbf e}_\Gamma$ (half the signed linewidth
contrast between its two eigenpolarizations). A constant polarization-basis rotation sends
$\hat{\mathbf e}_\Gamma$ to $\hat e_z$, after which the traceless part of Eq.~(\ref{eq:Jones}) is
Eq.~(\ref{eq:Hloc}), with $\dvec=h\mathbf S_0$ and $\gamma=\delta\Gamma$. This operational
parametrization does not predict how $\mathbf S_0$ is formed or prepared; it asks how the measured
doublet responds when differential loss is added. Common frequency and linewidth shifts
change neither the eigenvectors nor the EP condition. The local-response description
applies to weakly coupled polarization cells or to a local-density regime in which transverse
diffraction and TE--TM mixing across one resolution volume are perturbative. Residual nonlocal
coupling will round or displace the ideal contour and then requires a global cavity-mode analysis.
The present construction is not a Gross--Pitaevskii or Bogoliubov reduction. The polarization-basis
rotation is given in App.~\ref{app:polariton}.

\begin{figure}[!ht]
\centering
\includegraphics[width=0.99\linewidth]{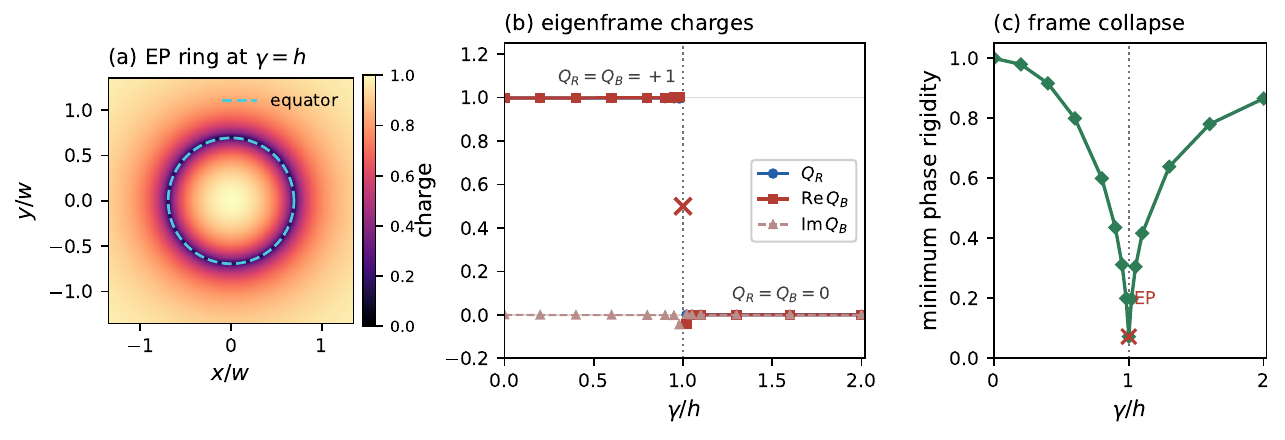}
\caption{\label{fig:core}
Real-space exceptional ring and eigenframe-charge transition. (a) At $\gamma=h$ the phase
rigidity vanishes on the skyrmion equator (dashed). (b) With the band fixed by spatial continuation,
$\QR$ and $\QB$ are $+1$ below the ring and $0$ above it; both are undefined on the defective
texture (cross). (c) The minimum phase rigidity collapses at the same threshold. The continuum
rigidity vanishes at the ring; the plotted minima are finite-grid samples. The charge sweep uses a
square of side length $192$, grid spacing $0.5$ ($384\times384$ points), $w=14$, and $h=1$
(App.~\ref{app:num}).
}
\end{figure}

The eigenvalues of the traceless generator are
\begin{equation}
\label{eq:eig}
\varepsilon^2 = |\dvec|^2 - \gamma^2 + 2i\gamma\,d_z ,
\end{equation}
so the two bands coalesce at an exceptional point where the discriminant vanishes and the traceless
part of the generator is nonzero, i.e.\ where
\begin{equation}
\label{eq:EP}
d_z = 0 \quad\text{and}\quad d_x^2 + d_y^2 = \gamma^2 .
\end{equation}
On a constant-modulus skyrmion the condition $d_z=0$ is met on the equator, and there
$d_x^2+d_y^2=h^2$. The whole equator is therefore exceptional at $\gamma=h$. This is a
spatially resolved EP ring inside one real-space eigenstate texture [Fig.~\ref{fig:core}(a)]. It is
distinct from the global coalescence of two confined optical-skyrmion modes studied in
Ref.~\cite{krol2023}, and from momentum-space exceptional rings in band theory.

The behavior of the biorthogonal charge across this ring is already visible in the equatorial
generator $H=h\,\sigma_x+i\gamma\,\sigma_z$. For the branch that is the lower band at
$\gamma<h$, the biorthogonal Bloch vector is
\begin{equation}
\label{eq:m2}
\mvec = \Bigl(\tfrac{-h}{\sqrt{h^2-\gamma^2}},\; 0,\; \tfrac{-i\gamma}{\sqrt{h^2-\gamma^2}}\Bigr),
\end{equation}
with $m_z$ purely imaginary and $\mvec\to\infty$ as $\gamma\to h^-$. Beyond the EP, this branch is
defined by spatial continuation rather than by an ordering of the imaginary eigenvalues. The phase rigidity
\begin{equation}
\label{eq:rig}
r = \frac{|\overlap|}{\sqrt{\langle \zL|\zL\rangle\,\langle \zR|\zR\rangle}}
\end{equation}
tends to zero at the same point: the two right eigenvectors coalesce, as do the two left
eigenvectors, and the coalesced state is self-orthogonal under the biorthogonal pairing. The
biorthogonal Bloch vector therefore becomes singular on the equatorial ring at the exceptional
point: its components diverge while the bilinear identity $m\cdot m=1$ --- which follows from the
completeness of the Pauli matrices and holds at every nondefective point --- continues to hold on
approach to the ring. The biorthogonal frame is defective at the ring, while the Bloch field stays
on the noncompact $\Sigmacom$ away from it. Equation~(\ref{eq:m2}) and the rigidity statement are
verified symbolically in App.~\ref{app:twolevel}.

We compute both eigenframe charges from the local right and left eigenvectors of
Eq.~(\ref{eq:Hloc}), with the band fixed by spatial continuation. Below the EP, $\QR$ and $\QB$
remain at $+1$ within a discretization error that decreases under fixed-box grid refinement
[Fig.~\ref{fig:core}(b)]. At $\gamma=h$, $\mvec$ diverges on the equator, phase rigidity
collapses, and no globally consistent band assignment exists. Immediately above the ring, a smooth
band exists again, and its class is fixed analytically, independently of any labeling convention.
Deform the generator to $H_t=(t\,d_x,t\,d_y,d_z)\cdot\sigmavec+i\gamma\sigma_z$ with $t:1\to0$;
its exceptional set requires $d_z=0$ and $t^2(d_x^2+d_y^2)=\gamma^2$. For $\gamma>h$ the second
equality is unreachable, since $d_x^2+d_y^2\le|\dvec|^2=h^2<\gamma^2$: the family is a homotopy of
nonsingular eigenframes, ending at $(d_z+i\gamma)\sigma_z$, whose two eigenlines are the constant
$\sigma_z$ eigenstates with charge $0$. The class is constant along a nonsingular homotopy, so both
eigenframe charges vanish throughout $\gamma>h$. For $\gamma<h$ the same auxiliary deformation
instead meets the exceptional set at $t=\gamma/h$ on the equator: in the extended $(t,\gamma)$
parameter space, the contraction is blocked by the same equatorial EP locus. The physical ring
therefore mediates a parameter-driven spectral eigenframe transition $+1\to0$ whose final value is
a property of the family, not of the continuation algorithm. A pointwise
principal square root instead swaps bands across the equator above the EP and produces
branch-dependent nonintegers; the spatial-continuation test, which recovers the analytic value
$0$, is recorded in App.~\ref{app:num}.

The two statements about protection are thus compatible. Homotopy protects the charge $Q$ of a
smooth trajectory. The plotted $\QR$ and $\QB$ characterize an instantaneous spectral frame. That
frame changes class only after becoming defective, and is smooth again on the other side of the
ring. Their equality follows from the eigenline-bundle argument in Sec.~\ref{sec:charges}; the
separate right- and left-field calculations cross-check it numerically.

\FloatBarrier

\section{Geometry of the exceptional set}
\label{sec:geometry}

The closed ring is not the generic exceptional set of a two-dimensional two-level problem. In an
arbitrary fixed polarization basis the traceless part of Eq.~(\ref{eq:Jones}) is
$\dvec\cdot\sigmavec+i\delta\Gamma\hat{\mathbf e}_\Gamma\cdot\sigmavec$. Its discriminant gives
the two basis-independent EP conditions
\begin{equation}
\dvec\cdot\hat{\mathbf e}_\Gamma=0,
\qquad |\dvec|^2=\delta\Gamma^2.
\label{eq:generalEP}
\end{equation}
They describe a defective EP provided $(\dvec,\delta\Gamma)\ne(\mathbf0,0)$; the excluded scalar
matrix is an ordinary twofold degeneracy. Equation~(\ref{eq:generalEP}) imposes two real conditions,
so its solutions in a two-dimensional texture are normally isolated. A contour requires an
additional constraint that makes the linewidth-matching condition hold along the first condition's
entire zero set.

\begin{figure}[!ht]
\centering
\includegraphics[width=0.99\linewidth]{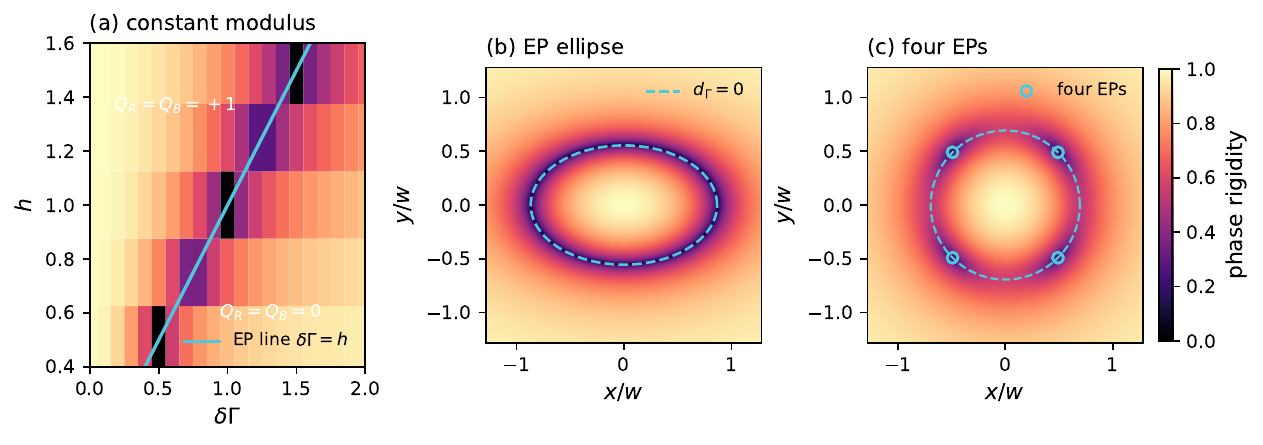}
\caption{\label{fig:geometry}
Geometry of the real-space exceptional set. (a) Minimum phase rigidity from raw equatorial
$H,H^\dagger$ diagonalization for constant-modulus textures. The analytic line
$\delta\Gamma=h$ separates the two EP-free eigenframe classes and is itself defective. (b) A
coordinate deformation at fixed $|\dvec|$ produces an elliptical EP
contour (dashed). (c) The smooth splitting anisotropy of Eq.~(\ref{eq:anisotropy}), with
$\epsilon=0.18$ and $\delta\Gamma=h$, leaves four isolated EPs (circles). In (b,c), color is the
finite-grid local phase rigidity, the circles mark the analytic EP positions, and the dashed curve
is $\dvec\cdot\hat{\mathbf e}_\Gamma=0$.
}
\end{figure}

For the constant-modulus family this constraint is explicit. The first condition selects the
equator and $|\dvec|=h$ there, so the whole equator becomes defective on the line
$\delta\Gamma=h$ [Fig.~\ref{fig:geometry}(a)]. The two EP-free sides carry the integer classes
$+1$ and $0$. A coordinate deformation
$\rho^2=(x/a_x)^2+(y/a_y)^2$ retains $|\dvec|=h$ and turns the circle into an ellipse, without
removing the charge transition [Fig.~\ref{fig:geometry}(b)]. Fixed-box refinement of this control
is given in App.~\ref{app:polariton}.

Amplitude anisotropy behaves differently. Consider the smooth splitting
\begin{equation}
h(r,\theta)=h[1+\epsilon g(r)\cos2\theta],\qquad
g(r)=\left(\frac{r}{r_{\rm eq}}\right)^2
\exp\!\left[1-\left(\frac{r}{r_{\rm eq}}\right)^2\right],
\label{eq:anisotropy}
\end{equation}
where $r_{\rm eq}=w\ln2$. The envelope vanishes at the core and at infinity and equals one on the
equator. At uniform $\delta\Gamma=h$, Eq.~(\ref{eq:generalEP}) then holds only at
$\cos2\theta=0$, leaving four EPs [Fig.~\ref{fig:geometry}(c)]. Low phase rigidity may persist on
the rest of the equator, but those points are not defective.

The distinction is therefore between coordinate anisotropy, which deforms a designed contour, and
splitting anisotropy, which removes the constraint that sustained it. We refer to the closed EP
locus as symmetry- or design-protected, not as a generic exceptional ring.

\FloatBarrier

\section{Localized differential loss and band obstruction}
\label{sec:localized}

\begin{figure}[!ht]
\centering
\includegraphics[width=0.99\linewidth]{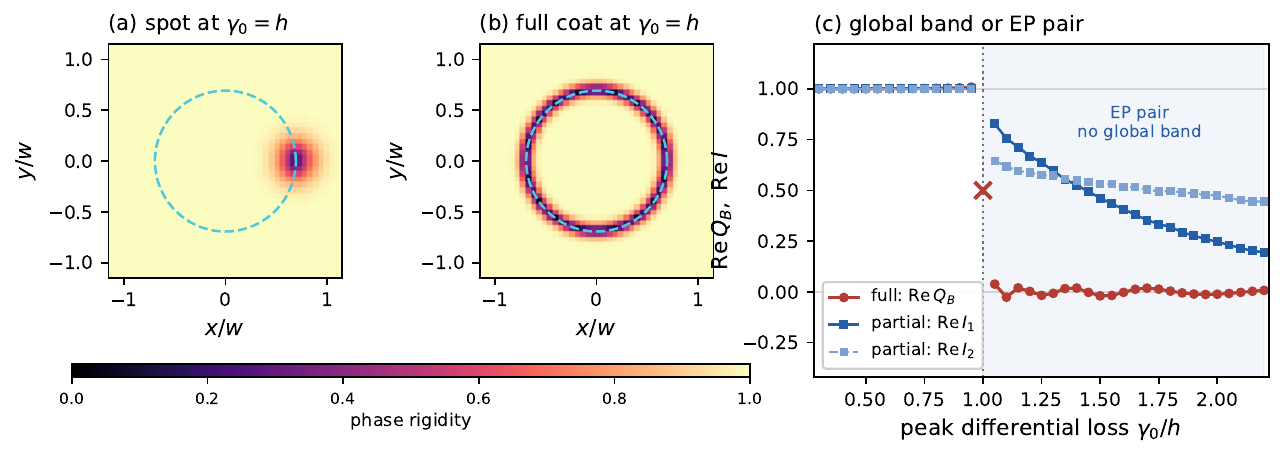}
\caption{\label{fig:loss}
Localized versus equatorial differential loss. (a) A narrow Gaussian spot at $\gamma_0=h$
touches the EP condition at one equatorial point. (b) A full equatorial coat makes the whole
equator defective. (c) The full coat is EP-free off threshold and its $\QB$ executes a quantized
$+1\to0$ transition. In the window~(\ref{eq:pairwindow}), a partial coat pins an EP pair; two
continuation trees then give different $\operatorname{Re}I_{\rm tree\,1,2}$, neither of which is a
topological charge. Panel (c)
uses a broader partial coat ($\sigma_s=12$) than the spot map in (a) ($\sigma_s=3$) to resolve the
two continuation paths. Its plotted range $1<\gamma_0/h\le2.2$ lies below the second threshold
$\exp(2r_{\rm eq}^2/\sigma_s^2)=3.70$; the numerical parameters are $w=14$ and $h=1$.
}
\end{figure}

The coverage of the prospective contour provides a second control. Let the differential loss be
either an azimuthally uniform Gaussian coat or a localized spot,
\begin{align}
\gamma_{\rm ring}(r)&=\gamma_0
\exp[-(r-r_{\rm eq})^2/(2\sigma_r^2)],\\
\gamma_{\rm spot}(x,y)&=\gamma_0
\exp[-((x-r_{\rm eq})^2+y^2)/(2\sigma_s^2)].
\label{eq:lossprofiles}
\end{align}
For the full-coat sweep below we use $\sigma_r=1$.
At $\gamma_0=h$, the full coat satisfies both EP conditions along the whole equator, whereas the
spot reaches them only at its center [Fig.~\ref{fig:loss}(a,b)]. Thus a small lossy area need not
produce a short EP arc: away from the exact touching point it produces only a neighborhood of low
phase rigidity.

The topology on the two sides depends on coverage. For the full coat, every texture with
$\gamma_0\ne h$ is EP-free and admits a global band label; its $\QB$ makes the same quantized
$+1\to0$ transition as the uniform model. Along the equator the Gaussian spot has minimum value
$\gamma_0\exp(-2r_{\rm eq}^2/\sigma_s^2)$. It therefore leaves an EP pair only in the localized
window
\begin{equation}
h<\gamma_0<h\exp(2r_{\rm eq}^2/\sigma_s^2),
\label{eq:pairwindow}
\end{equation}
or, more generally, when its equatorial profile takes values on both sides of $h$. At the upper
endpoint the pair meets at the antipode and disappears; a global band can then return.

The band obstruction in the window~(\ref{eq:pairwindow}) follows locally from the discriminant,
not from the continuation algorithm. Let $s$ and $u$ be tangent and normal coordinates at either
EP. For a transverse crossing of both conditions in Eq.~(\ref{eq:generalEP}),
\begin{equation}
\varepsilon^2=A\,\delta s+(C+iB)\,\delta u+O(|\delta\rvec|^2),\qquad AB\ne0,
\label{eq:localwinding}
\end{equation}
where $A,B,C$ are real. The Jacobian from $(\delta s,\delta u)$ to
$(\operatorname{Re}\varepsilon^2,\operatorname{Im}\varepsilon^2)$ has determinant $AB$. A small
loop, positively oriented in the $(s,u)$ frame, therefore gives the discriminant winding
$\operatorname{sgn}(AB)=\pm1$, so its square root changes sign. The two members of the pair have opposite windings. This monodromy obstructs a global
band assignment, and $\QR,\QB$ are not defined for that texture as global eigenframe charges.

Figure~\ref{fig:loss}(c) makes this obstruction numerical. The two continuation trees agree below
threshold, where a smooth band exists, and disagree above threshold, where the EP pair is present.
Their post-threshold complex integrals are diagnostics of continuation dependence, not two values
of $\QB$; the figure displays their real parts. The controlling variable is coverage of the
equatorial zero set in
Eq.~(\ref{eq:generalEP}), not the total area exposed to loss.

\FloatBarrier

\section{Polarimetric reconstruction and experimental scope}
\label{sec:protocol}

Under the local-response assumptions of Sec.~\ref{sec:ring}, spatially, spectrally, and
polarization-resolved cavity emission can be fit locally to two complex poles and two
right-eigenstate Stokes vectors $\mathbf S^R_\pm$. The EP set is the locus where the fitted poles
and their polarizations coalesce and the phase rigidity vanishes.
For a two-band generator, biorthogonality fixes the left-state Stokes vectors from the opposite
right band~\cite{hu2024},
\begin{equation}
\mathbf S^L_\pm=-\mathbf S^R_\mp.
\label{eq:leftStokes}
\end{equation}
The reconstructed left and right spinors give $m_a$, the local biorthogonal charge density, and the
phase rigidity. The explicit gauge construction and a raw-matrix cross-check are recorded in
App.~\ref{app:polariton}.

Ordinary Stokes tomography of one band gives $\QR$. On an EP-free compactified texture this also
fixes the integer $\QB=\QR$, although it does not determine the local biorthogonal field. Both bands
are required to reconstruct $\mvec$, phase rigidity, and the biorthogonal charge density. At an EP
contour or EP pair, no global eigenframe charge should be assigned.

The scales involved provide at least an order-of-magnitude reference. The EP condition is linewidth
matching, $\delta\Gamma\sim h$. In the phenomenological model of Ref.~\cite{hu2025} for a
liquid-crystal anisotropic microcavity, the coefficients are $\delta\Gamma=0.12$~meV and a Zeeman
coefficient $\Delta=0.15$~meV (band splitting $2\Delta$). Experimentally, Voigt exceptional points
and polarization Stokes textures have been measured in an anisotropic planar
microcavity~\cite{richter2019}. Separately, confined optical-skyrmion modes have been taken through
a modal EP in numerical work~\cite{krol2023}. For scale, the common cavity linewidth of the GaAs
platform of Ref.~\cite{cilibrizzi2016} corresponds to a Hamiltonian imaginary part
$\hbar/(2\tau_p)\approx0.09$~meV ($\tau_p=3.8$~ps); the differential contrast $\delta\Gamma$ is a
distinct quantity that must be independently measured or engineered. Among the momentum-dependent
terms neglected in the local fit, the leading polarization-mixing contribution is the TE--TM
splitting; with a coefficient of order $10\,\mu{\rm eV}\,\mu{\rm m}^2$~\cite{cilibrizzi2016}, it
contributes of order $\beta_{\rm LT}/w^2$
across a texture of width $w$, which is below the $0.1$~meV scale for $w\gtrsim1\,\mu$m, while
pseudospin textures in these platforms span tens of micrometres~\cite{cilibrizzi2016}. This
estimate covers only the TE--TM term and does not by itself establish the local-doublet limit: no
single existing device has been shown to combine a constant-modulus real-space texture, a tunable
differential loss, and the isolated-doublet regime. Whether a given cavity realizes a closed
contour, a displaced contour, or isolated exceptional points is decided by its global mode
structure, not by the local estimate.

This protocol assumes an isolated polarization doublet, spatial resolution sufficient for a local
fit, and nonlocal transverse coupling below the relevant linewidth resolution. It is a
local Jones or coupled-mode parametrization of a measured response, not a derivation of the global
cavity spectrum or a prediction of how the input Stokes texture is prepared. Nonlinear reservoir
dynamics, condensate preparation, and Bogoliubov excitations are outside its scope.

\section{Discussion}
\label{sec:discussion}

The result separates two uses of ``topological protection.'' The degree of a smooth trajectory is
unchanged because the normalized state remains a continuous map to $S^2$ with fixed boundary. A
local spectral frame also carries an integer class while it is nonsingular, but the frame may
switch classes as a control parameter passes through an EP. The real-space exceptional contour is
the singular set that mediates this parameter-driven spectral transition. It is not a time-resolved
trajectory through the EP and does not change the charge of the evolving condensate.

The novelty is not the occurrence of an EP in an optical skyrmion. Confined skyrmion modes have
already been taken through a global modal EP~\cite{krol2023}. Here one texture is diagonalized
point by point: at the transition its real-space equator is defective while the rest remains
nonsingular, and the integer eigenframe class is computed on both sides. The geometry analysis also states the limitation: without a
constant-splitting or equivalent design constraint, the contour resolves into isolated EPs.
On either EP-free side, $\QR=\QB$ is the Chern class of one eigenline rather than evidence for two
independent topological invariants. Biorthogonality supplies the local field, charge density, and
phase-rigidity diagnostic that become singular at the contour.

The content is therefore three concrete statements rather than one new invariant. First, an
object: a spatially resolved exceptional contour inside a single real-space texture, obtained by
diagonalizing one texture point by point, with the integer eigenframe class computed on both of
its EP-free sides---distinct from the global modal EP of confined skyrmion
modes~\cite{krol2023} and from momentum-space exceptional rings. Second, a criterion: the two
real EP conditions of Eq.~(\ref{eq:generalEP}) generically meet at isolated points in two
dimensions, so a closed contour exists only under a constant-splitting or equivalent design
constraint; coordinate anisotropy deforms the contour, whereas splitting anisotropy dissolves it
into isolated EPs. Third, a falsifiable dichotomy: a full equatorial coat realizes the quantized
$+1\to0$ transition, while a partial coat pins an EP pair inside the window of
Eq.~(\ref{eq:pairwindow}) and removes the global band label altogether---a distinction directly
accessible to the polarimetric protocol of Sec.~\ref{sec:protocol}.

For an isolated polarization doublet, the needed observables are the local complex resonances and
the Stokes textures of the two bands. This makes the distinction between a closed coating and a
partial coating directly testable. It does not establish a full driven-dissipative condensate
model, nor does it address nonlinear stability or mode competition.

The explicit topological calculation is for CP$^1$. The projector identity in
Sec.~\ref{sec:right} extends to a normalized CP$^{N-1}$ state, but the eigenframe charge and its
exceptional set require a separate higher-rank treatment.

\section*{Acknowledgements}

\paragraph{Funding information}
National High-Level Overseas Talent Program (KS21400126), Surface and Interface Synthetic
Chemistry (ZXP2025057), Jiangsu Distinguished Professorship Fund (SR21400225), and Research
Start-up Fund (NH21400525).

\paragraph{Data availability}
The numerical data generated in the high-performance-computing calculations, together with the
eigenstate-texture sweep code and figure-generating scripts, are available at Zenodo,
\url{https://doi.org/10.5281/zenodo.20862932}. The symbolic derivations are given in the Appendix.

\begin{appendix}

\section{Projective flow and the Gilbert term}
\label{app:gilbert}

We verify Eq.~(\ref{eq:gilbert}) symbolically. Parametrize the normalized spinor by Bloch angles,
$\hat z=(\cos\tfrac\theta2,\,e^{i\phi}\sin\tfrac\theta2)^{\top}$, so that
$n=(\sin\theta\cos\phi,\sin\theta\sin\phi,\cos\theta)$ and the projector $P=\hat z\hat z^\dagger$
satisfies $P^2=P$, $\mathrm{Tr}\,P=1$. Write the Hermitian part as
$H_0=b_0\mathbb 1+\mathbf b\cdot\sigmavec$ and the gain/loss coefficient as
$\Gamma=\gamma\sigma_z$. The trace-preserving projective evolution is
\[
\dot P=-i[H_0,P]+\{\Gamma,P\}-2\,\mathrm{Tr}(\Gamma P)\,P.
\]
The commutator gives $2\mathbf b\times\nvec$. The gain/loss contribution is
\[
\dot n_a|_\Gamma=\mathrm{Tr}\!\left[\sigma_a
\bigl(\{\Gamma,P\}-2\mathrm{Tr}(\Gamma P)P\bigr)\right],
\]
which gives
\[
\dot{\nvec}|_\Gamma = \bigl(-\gamma\sin 2\theta\cos\phi,\;
-\gamma\sin 2\theta\sin\phi,\;2\gamma\sin^2\theta\bigr),
\]
which equals $2\gamma(\Hzeff-n_z\nvec)$ identically and is tangent to $S^2$. The precession term is
tangent as well. Comparing the gain/loss term with
$\Heff-(\nvec\cdot\Heff)\nvec$, $\Heff=2\gamma\Hzeff$, shows the Gilbert equality term by term.
The symbolic check (SymPy) returns zero for the difference between the full right-hand side of
Eq.~(\ref{eq:gilbert}) and the projector calculation, and for $\nvec\cdot\dot{\nvec}$.

\section{Biorthogonal two-level model and the exceptional point}
\label{app:twolevel}

For $H=h\,\sigma_x+i\gamma\,\sigma_z$ we diagonalize $H=PDP^{-1}$; the columns of $P$ are the right
eigenvectors $|R_n\rangle$ and the rows of $P^{-1}$ are the left eigenvectors $\langle L_n|$,
automatically biorthonormal, $\langle L_m|R_n\rangle=\delta_{mn}$. The eigenvalues are
$E_\pm=\pm\sqrt{h^2-\gamma^2}$, real for $h>\gamma$ (unbroken $\PTsym$) and imaginary for
$h<\gamma$, coalescing at the exceptional point $\gamma=h$. For the branch denoted $-$ below the
EP and continued spatially above it, the biorthogonal Bloch vector
$m_a=\langle L_-|\sigma_a|R_-\rangle$ evaluates to
\[
\mvec = \left(\frac{-h}{\sqrt{h^2-\gamma^2}},\; 0,\; \frac{-i\gamma}{\sqrt{h^2-\gamma^2}}\right),
\]
so $m_z$ is purely imaginary and $|\mvec|\to\infty$ as $\gamma\to h^-$, even though
$\mvec\cdot\mvec=1$ throughout. The phase rigidity
$r=|\langle L_-|R_-\rangle|/\sqrt{\langle L_-|L_-\rangle\langle R_-|R_-\rangle}$
is invariant under reciprocal rescaling of the two eigenvectors and satisfies $r\to 0$ as
$\gamma\to h$, the signature of a defective coalescence.
These statements are confirmed symbolically (SymPy): $\mathrm{Im}\,m_z\neq 0$, and
$\lim_{h\to\gamma}|m_z|=\infty$.

\medskip
\noindent\textbf{Quantization on EP-free textures.}
The two-form $\Omega=(8\pi)^{-1}\epsilon_{abc}m_a\,\mathrm d m_b\wedge\mathrm d m_c$ is closed on
$\Sigmacom$: $\mathrm d\Omega$ is a holomorphic three-form on a complex surface and hence vanishes.
Since $\Sigmacom\cong T^*S^2$ deformation-retracts to $S^2$, and $\Omega$ restricts to the normalized
area form on the zero section, its integral over a texture that extends to the compactified plane is
the integer degree. Noncompactness does not alter this conclusion for a continuous map from a
compact domain. It matters only when the image escapes every compact subset, as it does when the
phase rigidity vanishes and $|\mvec|\to\infty$ at an EP. With separately Dirac-normalized left and
right states their overlap vanishes; in the biorthonormal gauge used above the overlap stays one
while the eigenvector norms and the spectral projector diverge.

Equivalently, $(\mathbb 1+\mvec\cdot\sigmavec)/2$ is the rank-one spectral projector onto the right
eigenstate line. Its Chern number is $\QB$. Orthogonal projection onto the same right line gives
$\QR$, so the two integer charges agree on every EP-free compactified texture. The two-band protocol
is needed for the local biorthogonal field and phase rigidity, not for this equality of integers.

A useful numerical trap is a purported ``smooth complex tilt''
$\mvec=\cosh a\,\nvec+i\sinh a\,\tvec$, where $\tvec$ is a unit tangent field along a skyrmion. If
$a(0)\neq0$, $\tvec$ winds at the core and the construction has a hidden point defect; its
noninteger integral measures that defect. Setting $a(0)=0$ removes it and restores the integer,
with the discretization error decreasing under grid refinement. The eigenstate texture gives
$\QB=+1$ below the ring and $0$ above it when the band is labeled continuously in space.

\section{Numerical method for the texture sweep}
\label{app:num}

We build the hedgehog field $\dvec=h(\sin f\cos\theta,\sin f\sin\theta,\cos f)$ on a square grid,
with $f(r)=\pi e^{-r/w}$. At each point we diagonalize
$H=\dvec\cdot\sigmavec+i\gamma\sigma_z$ and assemble $\nvec$ and $\mvec$ from the raw right and left
eigenvectors. The square has side length $192$ and grid spacing $0.5$, giving $384\times384$
points; $w=14$ and $h=1$. Fixed-box refinements converge to both integer plateaus.

For $\gamma\neq h$ the texture is nondegenerate over the simply connected plane, so a globally
smooth band label exists. We obtain it by spatial continuation, choosing between the two signs of
$\Ecal$ at each link to minimize the change from the previously assigned point. Above the ring,
the pointwise principal root is discontinuous where $d_z$ changes sign and silently swaps the two
bands. Near the threshold its apparent noninteger $\QB$ can grow under refinement; farther above
the threshold it may instead converge to a wrong branch-dependent noninteger. Spatial continuation
gives $\QB=0$ and leaves no sign-mismatched links away from the EP. Two independent continuation
trees agree for every EP-free control.

For the full equatorial coat of Sec.~\ref{sec:localized}, fixed-box refinement gives
\begin{center}
\begin{tabular}{c|cc}
grid spacing & $\QB(0.95h)$ & $\QB(1.20h)$\\
\hline
$0.75$  & $1.05727+0.03647i$ & $-0.04220+0.09930i$\\
$0.5$   & $1.03197-0.01159i$ & $-0.11926+0.04683i$\\
$0.375$ & $1.01747+0.00475i$ & $-0.05865-0.01334i$\\
$0.25$  & $1.00865+0.00048i$ & $0.00243+0.00463i$
\end{tabular}
\end{center}
so both plateaus converge to the integers shown in Fig.~\ref{fig:loss}(c). We also evaluated the
raw discriminant around small loops enclosing the two partial-coat EPs at $\gamma_0/h=1.5$; their
windings are $+1$ and $-1$. Finite differences of the same raw discriminant give nonzero local
Jacobians with determinants $AB$ of opposite sign, independently confirming
Eq.~(\ref{eq:localwinding}) and the square-root band exchange.

The exceptional transition is therefore a property of the instantaneous eigenstructure, not of a
single time-evolved right state. This distinction is also why the conserved texture charge $Q$ and
the jumping eigenframe charges in Fig.~\ref{fig:core} do not conflict.

\section{Jones basis rotation, tomography, and numerical controls}
\label{app:polariton}

For an isolated polarization doublet, the most general local Jones generator relevant here is
Eq.~(\ref{eq:Jones}). A common complex shift
$[\bar\omega-i\bar\Gamma]\mathbb 1$ changes the two resonance frequencies equally and leaves the
eigenvectors, phase rigidity, and EP condition unchanged. A constant SU(2) rotation aligns the
differential-loss axis with $\hat e_z$. The Pauli vector rotates by its associated SO(3) matrix, so
the traceless generator becomes $\dvec\cdot\sigmavec+i\delta\Gamma\sigma_z$, with
$\dvec=h\mathbf S_0$. The discriminant is invariant under this rotation and gives
Eq.~(\ref{eq:generalEP}) in the original basis.

Spatial and frequency-resolved polarimetry measures the two right-eigenstate Stokes vectors
$\mathbf S^R_\pm$. Equation~(\ref{eq:leftStokes}) then supplies the left-state Stokes vectors.
Writing $\theta^R_\pm=\arccos S^R_{z,\pm}$ and
$\phi^R_\pm=\operatorname{atan2}(S^R_{y,\pm},S^R_{x,\pm})$, one convenient gauge is
\begin{align}
|R_\pm\rangle&=\begin{pmatrix}e^{-i\phi^R_\pm}\cos(\theta^R_\pm/2)\\
\sin(\theta^R_\pm/2)\end{pmatrix},\\
\langle L_\pm|&\propto\begin{pmatrix}-e^{i\phi^R_\mp}\sin(\theta^R_\mp/2)&
\cos(\theta^R_\mp/2)\end{pmatrix}.
\end{align}
These spinors determine $m_a$, phase rigidity, and the local biorthogonal charge density. Their
direct reconstruction requires both bands. If only the integer class is sought on an EP-free
texture, however, $\QB=\QR$ and one-band tomography suffices.

As an external numerical check, we independently diagonalized 16,384 raw matrices $H_J$ and
$H_J^\dagger$, rather than evaluating a closed-form reduction. The right- and left-eigenvector
residuals are below $10^{-12}$. The Stokes antipode relation and the reconstructed $\mvec$ agree
with the direct left eigenvectors to the same tolerance.

For the elliptical control in Fig.~\ref{fig:geometry}(b), the fixed-box charges converge as
\begin{center}
\begin{tabular}{c|cc|cc}
grid spacing & $\QR(0.95h)$ & $\QB(0.95h)$ & $\QR(1.05h)$ & $\QB(1.05h)$\\
\hline
$1$    & $0.9872$ & $1.0076+0.0200i$ & $0.00225$ & $0.02435+0.00500i$\\
$0.5$  & $0.9966$ & $1.0014-0.00335i$ & $0.00057$ & $-0.00198+0.00164i$\\
$0.25$ & $0.9991$ & $0.99989-0.00037i$ & $0.00014$ & $0.000014-0.000105i$
\end{tabular}
\end{center}
at fixed physical box, verifying the same $+1\to0$ transition. The $\delta\Gamma=0$ control gives
$\QR\simeq\QB\simeq1$, while a texture-free generator gives both charges zero. The smooth envelope
in Eq.~(\ref{eq:anisotropy}) vanishes at the core and at infinity, so the four-EP control introduces
no hidden point defect and remains compactifiable.

\end{appendix}

\bibliography{references}

\begin{thebibliography}{10}
\providecommand{\url}[1]{\texttt{#1}}
\providecommand{\urlprefix}{URL }
\expandafter\ifx\csname urlstyle\endcsname\relax
  \providecommand{\doi}[1]{doi:\discretionary{}{}{}#1}\else
  \providecommand{\doi}{doi:\discretionary{}{}{}\begingroup
  \urlstyle{rm}\Url}\fi
\providecommand{\eprint}[2][]{\url{#2}}

\bibitem{skyrme1962}
T.~H.~R. Skyrme,
\newblock \emph{A unified field theory of mesons and baryons},
\newblock Nuclear Physics \textbf{31}, 556 (1962).

\bibitem{nagaosa2013}
N.~Nagaosa and Y.~Tokura,
\newblock \emph{Topological properties and dynamics of magnetic skyrmions},
\newblock Nature Nanotechnology \textbf{8}, 899 (2013).

\bibitem{tsesses2018}
S.~Tsesses, E.~Ostrovsky, K.~Cohen, B.~Gjonaj, N.~H. Lindner and G.~Bartal,
\newblock \emph{Optical skyrmion lattice in evanescent electromagnetic fields},
\newblock Science \textbf{361}, 993 (2018),
\newblock \eprint{1805.11839}.

\bibitem{shen2024}
Y.~Shen, Q.~Zhang, P.~Shi, L.~Du, X.~Yuan and A.~V. Zayats,
\newblock \emph{Optical skyrmions and other topological quasiparticles of
  light},
\newblock Nature Photonics \textbf{18}, 15 (2024),
\newblock \eprint{2205.10329}.

\bibitem{cilibrizzi2016}
P.~Cilibrizzi, H.~Sigurdsson, T.~C.~H. Liew, H.~Ohadi, A.~Askitopoulos,
  S.~Brodbeck, C.~Schneider, I.~A. Shelykh, S.~H{\"o}fling, J.~Ruostekoski and
  P.~G. Lagoudakis,
\newblock \emph{Half-skyrmion spin textures in polariton microcavities},
\newblock Physical Review B \textbf{94}, 045315 (2016),
\newblock \eprint{1602.04711}.

\bibitem{heiss2012}
W.~D. Heiss,
\newblock \emph{The physics of exceptional points},
\newblock Journal of Physics A: Mathematical and Theoretical \textbf{45},
  444016 (2012).

\bibitem{ashida2020}
Y.~Ashida, Z.~Gong and M.~Ueda,
\newblock \emph{Non-{H}ermitian physics},
\newblock Advances in Physics \textbf{69}, 249 (2020),
\newblock \eprint{1910.00987}.

\bibitem{bergholtz2021}
E.~J. Bergholtz, J.~C. Budich and F.~K. Kunst,
\newblock \emph{Exceptional topology of non-{H}ermitian systems},
\newblock Reviews of Modern Physics \textbf{93}, 015005 (2021).

\bibitem{shen2018}
H.~Shen, B.~Zhen and L.~Fu,
\newblock \emph{Topological band theory for non-{H}ermitian {H}amiltonians},
\newblock Physical Review Letters \textbf{120}, 146402 (2018),
\newblock \eprint{1706.07435}.

\bibitem{kawabata2019}
K.~Kawabata, K.~Shiozaki, M.~Ueda and M.~Sato,
\newblock \emph{Symmetry and topology in non-{H}ermitian physics},
\newblock Physical Review X \textbf{9}, 041015 (2019),
\newblock \eprint{1812.09133}.

\bibitem{gong2018}
Z.~Gong, Y.~Ashida, K.~Kawabata, K.~Takasan, S.~Higashikawa and M.~Ueda,
\newblock \emph{Topological phases of non-{H}ermitian systems},
\newblock Physical Review X \textbf{8}, 031079 (2018),
\newblock \eprint{1802.07964}.

\bibitem{richter2019}
S.~Richter, H.-G. Zirnstein, J.~Z{\'u}{\~n}iga-P{\'e}rez, E.~Kr{\"u}ger,
  C.~Deparis, L.~Trefflich, C.~Sturm, B.~Rosenow, M.~Grundmann and
  R.~Schmidt-Grund,
\newblock \emph{Voigt exceptional points in an anisotropic {ZnO}-based planar
  microcavity: Square-root topology, polarization vortices, and circularity},
\newblock Physical Review Letters \textbf{123}, 227401 (2019),
\newblock \doi{10.1103/PhysRevLett.123.227401}.

\bibitem{krol2023}
X.~Luo, Y.~Cai, X.~Yue, W.~Lin, J.~Zhu, Y.~Zhang and F.~Li,
\newblock \emph{Non-{H}ermitian control of confined optical skyrmions in
  microcavities formed by photonic spin-orbit coupling},
\newblock Photonics Research \textbf{11}, 610 (2023),
\newblock \doi{10.1364/PRJ.478364}.

\bibitem{hu2025}
Y.-M.~R. Hu, E.~A. Ostrovskaya, A.~Yakimenko and E.~Estrecho,
\newblock \emph{Emergent momentum-space topological pseudospin defects in
  non-{H}ermitian systems},
\newblock Optics Express \textbf{34}, 13580 (2026),
\newblock \doi{10.1364/OE.591803},
\newblock \eprint{2509.14605}.

\bibitem{kunst2018}
F.~K. Kunst, E.~Edvardsson, J.~C. Budich and E.~J. Bergholtz,
\newblock \emph{Biorthogonal bulk-boundary correspondence in non-{H}ermitian
  systems},
\newblock Physical Review Letters \textbf{121}, 026808 (2018),
\newblock \eprint{1805.06492}.

\bibitem{edvardsson2020}
E.~Edvardsson, F.~K. Kunst, T.~Yoshida and E.~J. Bergholtz,
\newblock \emph{Phase transitions and generalized biorthogonal polarization in
  non-{H}ermitian systems},
\newblock Physical Review Research \textbf{2}, 043046 (2020).

\bibitem{mandal2024}
I.~Mandal,
\newblock \emph{Identifying gap-closings in open non-{H}ermitian systems by
  biorthogonal polarization},
\newblock Journal of Applied Physics \textbf{135}, 094402 (2024),
\newblock \eprint{2401.12213}.

\bibitem{hu2024}
Y.-M.~R. Hu, E.~A. Ostrovskaya and E.~Estrecho,
\newblock \emph{Generalized quantum geometric tensor in a non-{H}ermitian
  exciton-polariton system},
\newblock Optical Materials Express \textbf{14}, 664 (2024),
\newblock \eprint{2306.00351}.

\end{thebibliography}

\end{document}